\documentclass[preprintnumbers, floatfix, letterpaper, superscriptaddress,nofootinbib, twocolumn]{revtex4}
\usepackage{graphicx}
\usepackage{microtype}
\usepackage{amsmath}
\usepackage{amssymb}
\usepackage{subfigure}
\usepackage{hyperref}
\usepackage{url}
\usepackage{xcolor}
\usepackage{color}
\usepackage{mathrsfs}
\usepackage{calrsfs}
\usepackage{amsfonts}
\usepackage{latexsym}
\usepackage{ragged2e}
\usepackage{epsfig}
\usepackage{textcomp}
\usepackage{phaistos}
\makeatletter
\renewcommand\@makefnmark{\hbox{\@textsuperscript{\normalfont\color{purple}\@thefnmark}}}
\renewcommand\@makefntext[1]{%
  \parindent 1em\noindent
            \hb@xt@1.8em{%
                \hss\@textsuperscript{\normalfont\@thefnmark}}#1}
\makeatother

\usepackage{caption}
\DeclareCaptionJustification{justified}{\leftskip=0pt \rightskip=0pt \parfillskip=0pt plus 1fil}
\definecolor{vividviolet}{rgb}{0.62, 0.0, 1.0}
\definecolor{amaranth}{rgb}{0.9, 0.17, 0.31}
\definecolor{palatinateblue}{rgb}{0.15, 0.23, 0.89}
\definecolor{brightpink}{rgb}{1.0, 0.0, 0.5}
\definecolor{cornflowerblue}{rgb}{0.39, 0.58, 0.93}
\definecolor{deepcarminepink}{rgb}{0.94, 0.19, 0.22}
\definecolor{radicalred}{rgb}{1.0, 0.21, 0.37}

\hypersetup{ linktoc=all,
    colorlinks, linkcolor={palatinateblue},
    citecolor={brightpink}, urlcolor={amaranth}
}

\graphicspath{{Images/}}

\graphicspath{{Images/}}

\makeatletter
\def\@fnsymbol#1{\ensuremath{\ifcase#1\or $\textleaf$ \or $\PHplaneTree$
\else\@ctrerr\fi}}%

\makeatother

\def\sideremark#1{\ifvmode\leavevmode\fi\vadjust{\vbox to0pt{\vss
 \hbox to 0pt{\hskip\hsize\hskip1em
 \vbox{\hsize1.5cm\tiny\raggedright\pretolerance10000
 \noindent #1\hfill}\hss}\vbox to8pt{\vfil}\vss}}}%
\begin{document}

\title{Schwinger Pair Production and the Extended Uncertainty Principle: \\Can Heuristic Derivations Be Trusted?}

\author{Yen Chin \surname{Ong}}
\email{ycong@yzu.edu.cn}
\affiliation{Center for Gravitation and Cosmology, College of Physical Science and Technology, Yangzhou University, \\180 Siwangting Road, Yangzhou City, Jiangsu Province  225002, China}
\affiliation{School of Aeronautics and Astronautics, Shanghai Jiao Tong University, Shanghai 200240, China}

\begin{abstract}
The rate of Schwinger pair production due to an external electric field can be derived heuristically from the uncertainty principle. In the presence of a cosmological constant, it has been argued in the literature that the uncertainty principle receives a correction due to the background curvature, which is known as the ``extended uncertainty principle'' (EUP). We show that EUP does indeed lead to the correct result for Schwinger pair production rate in anti-de Sitter spacetime (the case for de Sitter spacetime is similar), provided that the EUP correction term is \emph{negative} (positive for the de Sitter case). We compare the results with previous works in the EUP literature, which are not all consistent. Our result further highlights an important issue in the literature of generalizations of the uncertainty principle: how much can heuristic derivations be trusted?
\begin{center}

\end{center}
\end{abstract}

\maketitle

\section{Schwinger Pair Production from the Uncertainty Principle}\label{1}

The quantum vacuum is teeming with virtual particles, whose fleeting existence is governed by the uncertainty principle. On the other hand, if we apply a sufficiently strong external electric field, we can ``boil the vacuum'' \cite{pc} and create real particle pairs from the virtual ones. This is the well known Schwinger effect \cite{Schwinger}. There are many ways to derive the Schwinger critical field and the corresponding pair production rate. However, a heuristic derivation can already give us some insights as to why such an effect should occur (in the Appendix we briefly discuss the Euclidean method).

Consider a virtual electron-positron pair in a constant electric field of strength\footnote{We shall work with the units in which $c=G=4\pi \epsilon_0 =1$ but $\hbar \neq 1$, so $\hbar$ has the dimension of area, while mass and charge have the dimension of length. The electric field has dimension of inverse length.} $E$. Suppose the particles move apart from each other by a distance $\ell$, then the amount of energy they receive from the electric field is $eE\ell$. The pair will become real if $eE\ell > 2m_e$, i.e., if the energy exceeds the rest mass of the two particles. The typical separation of the virtual pair is of the order of the Compton wavelenth $2\pi \hbar/m_e$. This can be derived from the Heisenberg uncertainty principle as follows. 
First, denote the characteristic length scale $\ell \sim \Delta x$, 
and\footnote{We assume that the speed $v$ is not too large to require relativistic correction for the momentum. In any case, for $v$ not too close to 1, the $\gamma$-factor is of order unity which can be neglected in our heuristic approach.} $\Delta p \sim m_e$. Then the uncertainty relation $\Delta x \Delta p \sim \hbar/2$ implies that $\ell \sim \hbar/(2m_e)$. This is the Compton wavelength $\ell_C=2\pi \hbar/m_e$ up to a dimensionless constant $4\pi$. Thus the condition that the virtual pair becomes real is the inequality
\begin{equation}\label{condition}
 e E\ell_C = 4\pi \ell eE > 2m_e, 
\end{equation}
which implies that the Schwinger critical field $E_S$ should satisfy (up to a constant $1/\pi$ factor), the relation
\begin{equation}
\frac{m_e^2}{\hbar e E} \sim 1,
\end{equation}
This is indeed the case. In conventional SI units, we have
\begin{equation}
E_s = \frac{m_e^2 c^3}{e\hbar} \approx 1.32 \times 10^{18} \text{V/m}.
\end{equation}
In other words, a heuristic argument that leads to Eq.(\ref{condition}) leads to a dimensionless quantity that governs the essential physics.
The Schwinger pair production rate, which we will denote as $\Gamma$, is proportional to $\exp\left[-S(E)\right]$, where
\begin{equation}\label{exp}
S(E)= \frac{\pi m_e^2}{\hbar eE},
\end{equation}
which is a constant multiple of the left hand side expression in Eq.(2). 

{So far, this heuristic, purely quantum mechanical ``derivation'' is a textbook material \cite{textbook}, which is of course more of a hindsight and consistency check.
One crucial step here involves putting $S(E)$ into an exponential. How should we understand this using only basic quantum mechanics?
Since we do not see copious production of electrons/positrons from the quantum vacuum in everyday life, such event is probably suppressed. That is to say, pair creation needs to overcome some kind of potential barrier. Quantum tunneling\footnote{In fact the Schwinger effect \emph{is} a tunneling process, as can be appreciated from more rigorous derivations, e.g. the instanton method \cite{0005078}.} probability is given by 
\begin{equation}
\mathcal{P} \sim \exp\left[-\frac{\ell p}{\hbar}\right],
\end{equation}
where $p\sim m_e$ and $\ell \sim m_e/eE$ from the above discussion. This explains the exponential probability in Eq.(\ref{exp}).
In any case, if we now accept such a heuristic argument, we can put it to use to ``derive'' pair-production rate when the uncertainty principle has been modified. 
}

In the following we will generalize this argument to derive the Schwinger effect in anti-de Sitter spacetime, in which according to the literature, uncertainty principle must be replaced by the ``extended uncertainty principle''. The result will turn out to be the same as the known formula in the literature, obtained via rigorous calculations. This suggests \emph{a posteriori} that the heuristic derivation has some merits. Nevertheless, for this heuristic derivation to work, the so-called ``EUP parameter'' needs to be of a different \emph{sign} compared to some earlier works in the literature. Indeed, the vast literature of generalizations of uncertainty principles to curved spacetimes and/or considering quantum gravitational effects, is full of various heuristic arguments, and the results are not all consistent with one another. We will discuss this discrepancies in details in this work. This raises the important issue that is nevertheless not resolved in this work: how much can heuristic derivations be trusted?

\section{Schwinger Pair Production in Anti-de Sitter Spacetime}\label{2}

The Schwinger pair production rate receives a correction in the presence of a nonzero cosmological constant, $\Lambda$. 
In this work we will focus on the anti-de Sitter (AdS) case, which corresponds to $\Lambda <0$ (the case for $\Lambda > 0$, i.e. in de Sitter (dS) spacetime, is similar, and will be discussed later). With $L$ denoting the curvature length scale of the AdS spacetime, the pair production rate is known from the literature to be $\exp\left[-S(E,L)\right]$, where\footnote{In the square root sign there appears an additional term inversely proportional to $L^4$ \cite{0501169, 0803.2555, 1407.4569}, which is related to the famous Breitenlohner-Freedman bound \cite{BF} in AdS spacetime, if one considers one-loop vacuum amplitude effect. This term does not appear in our work. Our expression is the same as, e.g. Eq.(2.28) of \cite{0501169}.}
\begin{flalign}
S(E,L)&:=2 \pi L^2 \hbar^{-1} \left(eE - \sqrt{(eE)^2-m_e^2/L^2}\right)\\ &\approx \frac{1}{\hbar}\left[\frac{\pi  m_e^2}{ eE} + \frac{1}{4}\frac{\pi m_e^4}{ e^3E^3 L^2}\right], \label{ads}
\end{flalign}
up to $1/L^2$ order in the large $L$ series expansion \cite{0501169, 0803.2555, 1804.04140}. A derivation using Euclidean method (Wick rotation) is provided in the Appendix.
Note that the pair production is suppressed compared to the Minkowski case. On the other hand, the rate will be enhanced in de Sitter spacetime (heuristically, positive cosmological constant that drives the expansion of the Universe also makes separating particle pairs easier; a negative cosmological constant acts in an opposite manner.)

The question we are interested in is this: \emph{can we derive Eq.(\ref{ads}) with a suitable correction to the uncertainty principle?} As we shall see, the answer is yes, but not without leaving a puzzle behind concerning the sign of the correction parameter. 

Such a correction to the uncertainty principle is known as the ``extended uncertainty principle'' (EUP), which takes the form
\begin{equation}\label{EUP}
\Delta x \Delta p \sim \frac{\hbar}{2}\left[1+\frac{\beta(\Delta x)^2}{L^2}\right].
\end{equation}
The parameter $\beta$ is often taken to be of order unity. 
There have been some debates concerning the \emph{sign} of $\beta$, an unresolved issue that we will discuss in the next section. For now, let us take Eq.(\ref{EUP}) for granted and repeat the calculation in Sec.(\ref{1}).

Eq.(\ref{EUP}) is a quadratic equation in $\Delta x$ and thus gives two possible solutions 
\begin{equation}\label{pm}
\Delta x_{\pm} = \frac{L(L\Delta p  \pm \sqrt{L^2 (\Delta p)^2 - \beta \hbar^2})}{\beta \hbar}.
\end{equation}
However, $\Delta x_+ \sim 2\Delta p L^2/(\beta \hbar) - \hbar/(2\Delta p) + \mathcal{O}(\beta/L^2)$ in large $L$ limit, which is divergent. Therefore $\Delta x_-$ is the only sensible solution that yields the correct limiting behavior: $\Delta x_- \sim \hbar/(2\Delta p) + \mathcal{O}(\beta/L^2)$. Thus, with $\Delta p \sim m_e$, we have
\begin{equation}
\ell \sim \Delta x_- \sim \frac{L(m_e L - \sqrt{L^2 m_e^2 - \beta \hbar^2})}{\beta \hbar}.
\end{equation}

From Eq.(\ref{condition}), one can obtain the modified Schwinger critical field condition:
\begin{flalign}
1 & \sim \frac{m_e \beta \hbar}{2\pi e E L}\left(m_e L - \sqrt{m_e^2L^2 - \beta \hbar^2}\right)^{-1} \\ 
& =\frac{\beta \hbar}{2\pi e EL^2}\left[\frac{2 m_e^2 L^2}{\beta \hbar^2}-\frac{1}{2} + \mathcal{O}\left(\frac{\beta}{L^2}\right)\right] \\ 
& = \frac{1}{\pi^2 \hbar}\left[\frac{\pi m_e^2}{eE}-\frac{\pi \hbar^2}{4}\frac{\beta}{eEL^2}\right] + \mathcal{O}\left(\frac{\beta^2}{L^4}\right).
\end{flalign}
Dropping the constant prefactor, the expression in the square bracket should be compared to the expression in the square bracket of Eq.(\ref{ads}).

Since the characteristic field strength is $E \sim m_e/(2\pi \ell e)$, we also have 
\begin{flalign}
-\frac{\pi \hbar^2}{4}\frac{\beta}{e EL^2} & \sim -\frac{\pi^2 \hbar}{2 m_e L}\left(m_e L - \sqrt{m_e^2L^2 - \beta \hbar^2}\right) \\ \notag
& = -\frac{1}{4}\frac{\beta \hbar^3 \pi^2}{m_e^2 L^2} + \mathcal{O}\left(\frac{\beta^2}{L^4}\right).
\end{flalign}
Therefore, up to the same order of the series expansion, $\hbar = m_e^2/(\pi e E)$. Consequently, we have
\begin{equation}
-\frac{\pi \hbar^2}{4}\frac{\beta}{e EL^2} = - \frac{m_e^4 \beta}{4\pi e^3 E^3L^2}.
\end{equation}
Comparing this with Eq.(\ref{ads}), we conclude that
\begin{equation}
\beta = -\pi^2.
\end{equation}
While the exact numerical value is probably not important in such a heuristic treatment anyway, we note that the sign of the EUP correction is \emph{negative}\footnote{From Eq.(\ref{pm}), one observes that the terms under the square root must be nonnegative, so $L^2(\Delta p)^2 > \beta\hbar^2$. If $\beta$ is negative, one might worry that we might have $\Delta p =0$. However, it is unlikely that EUP considered in Eq.(\ref{EUP}) is complete -- there are likely higher order correction terms in $\Delta x/L$, similar to those in the case of generalized uncertainty principle (GUP), see e.g. \cite{0204110}. We will discuss GUP below.}. This is a surprising curiosity. Let us now compare this result with other works in the literature. 

\section{The Sign of Extended Uncertainty Principle Parameter}\label{3}
Initially, EUP was motivated by Park from the point of view that such a form of the uncertainty principle would allow a heuristic derivation of the Hawking temperature of black holes in AdS or dS spacetimes \cite{0709.2307} (further analysis of black hole thermodynamics in this context was carried out in \cite{0411086v1}). For example, the Hawking temperature of a Schwarzschild-AdS spacetime in $d$-dimension is given by
\begin{equation}
T=\frac{\hbar\left[ (d-3)L^2+(d-1)r_+^2\right] }{4\pi L^2 r_+}.
\end{equation}
Consider a temperature of a typical photon emitted by the black hole (in the unit that the Boltzmann constant $k_B=1$), $T = E = pc$. From the EUP, we can indeed heuristically derive the correct form
\begin{equation}
T\sim \Delta p \sim \frac{\hbar}{2}\left[\frac{1}{\Delta x} + \frac{\beta (\Delta x)}{L^2}\right].
\end{equation}
Such technique is a direct generalization of the discussion in asymptotically flat case \cite{0106080}, in which $\Delta x \sim r_+$ is the horizon scale (a typical Hawking quanta can materialize in a ``quantum atmosphere'' that extends quite far away from the horizon, as emphasized by Giddings \cite{1511.08221}).
This ``derivation'' implies that $\beta$ is positive in AdS spacetime, and negative in dS case\footnote{In \cite{0709.2307}, the absolute value $\beta$ is also dimensional dependent, but because of the heuristic nature of the Hawking temperature derivation, it is not clear that the constant numerical coefficients involved should be taken too seriously. Therefore we shall just focus our discussion on the sign of $\beta$.}. 

Shortly after, Bambi and Urban \cite{Bambi} argued that contrary to Park's proposal, the sign for $\beta$ in de Sitter spacetime should be positive. More recently, Lake et al. proposed a derivation of EUP from superpositions of geometries \cite{1812.10045}, in which the sign of EUP parameter is positive by construction. More specifically, $\beta$ is proportional to the square of a standard deviation (called ``momentum space smearing scale'' and denoted $\tilde{\sigma}_g$ in \cite{1812.10045}).

That is to say, our result in this work neither agrees with Lake et al. \cite{1812.10045} nor with Park \cite{0709.2307} (also not with the anti-Snyder-de Sitter model of Mignemi \cite{1110.0201,1911.08921}). In fact, there are at least three distinct possibilities:
\begin{itemize}
\item[(1)] $\text{sgn}(\beta) = \text{sgn}(\Lambda)$
\item[(2)] $\text{sgn}(\beta) = -\text{sgn}(\Lambda)$ ~~~ (Park \cite{0709.2307})
\item[(3)] $\beta > 0$ ~~~ (Lake et al. \cite{1812.10045})
\end{itemize}
Actually in \cite{1812.10045} Lake et al. suggested that their model ``rules out the physical existence of anti-de Sitter space ($\Lambda < 0$), since $l_\text{dS} = \sqrt{3/\Lambda}$ and $\tilde{\sigma}_g \simeq \hbar{\sqrt{\Lambda/3}}$ are, of course, required to be real'', although one could also imagine setting $\tilde{\sigma}_g \simeq \hbar{\sqrt{-\Lambda/3}}$ for AdS space, of which $l_\text{AdS} = \sqrt{-3/\Lambda}$ (in our notation, $L \equiv l_\text{(A)dS}$). This would lead to option (3) above. We remark that Bambi and Urban \cite{Bambi} seem to correspond to option (3).
In this section, we will attempt to further strengthen the argument for the case $\text{sgn}(\beta)=\text{sgn}(\Lambda)$, though we cannot be confident that this is indeed the correct one; see the Discussion section.

First we note that there have been recent attempts to give EUP a more rigorous foundation from other points of view, see, e.g. \cite{F, 0306080, 1804.02551, Mignemi, 0911.5695}. Notably, EUP correction can be viewed as a classical curvature correction due to the underlying geometry \cite{1804.02551, 9405067}. This is different from GUP correction (see Eq.(\ref{GUP}) below) which is quantum gravitational in nature. 

On a similar note, in the 1960s, Judge essentially showed that on a unit circle $S^1$, the uncertainty principle should take the form\footnote{Judge was actually discussing an equivalent problem: the uncertainty principle between angular momentum $L_z$ and angle $\varphi$.}
\begin{equation}
\Delta x \Delta p \geqslant \frac{\hbar}{2}\left[1-C(\Delta x)^2\right],
\end{equation}
where $C$ is a constant, argued to be $3/\pi^2$ \cite{judge1, judge2}; see also \cite{1512.05716}. This suggests that the underlying \emph{topology} of the manifold might also be relevant for the form of the uncertainty principle. In particular for $S^1$, which is closed, the sign of the EUP parameter would be negative. In a later analysis by Lake et al., e.g. \cite{1912.07094}, the EUP term is obtained from a superposition of Euclidean geometries, which are also intrinsically flat like $S^1$, but of course with different topology than $S^1$, and the EUP term turns out to be positive.

Note that, on the other hand, \cite{1804.02551} gives good arguments that if the underlying spatial geometry is positively curved, then the corresponding EUP should have a negative correction term and conversely, a negatively curved spatial geometry should give rise to a positive correction term, at least when the geometry is of constant sectional curvature (This is in contrast with the discussion in the previous paragraph, in which the sign of the EUP term appears to be related to the topology of the background, rather than its geometry). This would suggest that de Sitter spacetime, whose global spatial section is $S^3$, should correspond to negative EUP parameter. Nevertheless, one has to keep in mind that both de Sitter and anti-de Sitter spacetimes are maximally symmetric, so one could always choose a foliation such that the spatial slices are either positively curved, flat, or negatively curved, so this argument is suggestive at best.

In fact, for locally asymptotically AdS spacetimes, it is well-known that there are topological black hole solutions. Their Hawking temperature takes the form \cite{9808032}
\begin{equation}
T=\frac{  \hbar\left[k(d-3)L^2+(d-1)r_+^2 \right]}{4\pi L^2 r_+},
\end{equation}
where $k=+1,0,-1$ correspond to horizons that are positively curved, flat, and negatively curved, respectively. The heuristic argument to derive Hawking temperature discussed above therefore only works for $k=1$ case, and even then such subtleties mean that it becomes rather doubtful whether the heuristic argument works as intended. Note that for the $k=0$ case, AdS toral or planar black hole has temperature that is directly proportional to $r_+$, not inversely proportional to it as in asymptotically flat spacetime. If some form of modified uncertainty principle exists that would allow us to derive Hawking temperature in the manner discussed above, then it must take the form $\Delta x /\Delta p = \text{const.}$, which is not the usual Heisenberg form plus a correction term. This would be rather surprising indeed as one can take both $\Delta x$ and $\Delta p$ to be arbitrarily small, while keeping their ratio constant. In other words, Hawking radiation of AdS black holes depends on the underlying topology, which does not seem easily encoded by just a single form of EUP. For a different criticism of  \cite{0709.2307}, see \cite{0607010}.

The Schwinger effect, on the other hand, is independent of $k$. This can be readily shown, for example, by deriving the particle production rate from Euclidean method (Wick rotation), as we show in the Appendix. Our heuristic derivation thus fixes the sign of EUP parameter in a more concrete, straightforward, manner. 

Our work is, in any case, not the first to employ EUP to derive Schwinger effect in the presence of a cosmological constant. Hamil and Merad had previously derived Schwinger effect in \emph{de Sitter} spacetime by employing a much more rigorous method than ours \cite{Hamil}. They solved EUP-modified Klein-Gordon equation and obtained the pair production rate, which is known from earlier literature \cite{0803.2555, 1401.4137} to be, up to the first correction term\footnote{The full expression of $S(E,L)$ in de-Sitter spacetime is (see, e.g. \cite{1411.1787})
\begin{equation}
S(E,L)_\text{dS-EUP}=2\pi L^2 \hbar^{-1}\left[\sqrt{(eE)^2+\left(m/L\right)^2}-eE\right],
\end{equation}
having ignored the term that corresponds to the one-loop effect mentioned in Footnote 2.
},
\begin{equation}\label{dseup}
\Gamma_\text{dS-EUP}=\frac{\pi m^2}{eE} - \frac{1}{4}\frac{\pi m^4}{e^3 E^3 L^2},
\end{equation}
which corresponds to $\beta > 0$ in our work, as expected. This seems strange at first since Hamil and Merad actually assumed from the beginning that EUP in de Sitter spacetime corresponds to $\beta < 0$ (in our notation). However, there appears to be a typo of a sign (going from Eq.(56) and Eq.(57) to Eq.(59) in their paper), which seems to indicate that in order to match Eq.(\ref{dseup}), they should have $\beta > 0$ instead. Nevertheless, much of the calculations in \cite{Hamil} needs to be repeated with $\beta > 0$ to see if this gives consistent results, as the corresponding equations are not readily obtained just by reversing a few signs.

We shall also remark that various authors have employed EUP with positive $\beta$ without specifying whether it corresponds to either dS or AdS (see, e.g., \cite{9311147, 9412167}), but based their motivations on the ground that this recovers the symmetry with the generalized uncertainty principle (GUP), which is a quantum gravitational correction to the Heisenberg uncertainty principle \cite{9301067,9305163,9904025,9904026,6,7,8}:
\begin{equation}\label{GUP}
\Delta x \Delta p \sim \frac{\hbar}{2}\left[1+\frac{\alpha(\Delta p)^2}{\hbar}\right],
\end{equation}
in which $\alpha$ is often taken to be positive. Indeed GUP with positive $\alpha$ can be derived from various means, including various quantum gravitational arguments (see also, \cite{1001.1205}). Curiously, even for the case of GUP, there are still some indications that $\alpha$ might be negative. For example, a lattice ``spacetime crystal'' gives rise to such a GUP \cite{0912.2253}. 
Negative GUP parameter is also needed if one accepts that Wick-rotation can be applied to obtain GUP-corrected black hole temperature from a Schwarzschild-like black hole with higher order terms \cite{1407.0113}. More recently, non-commutative geometry \cite{KLVY} and corpuscular gravity, were also shown to give rise to negative $\alpha$ \cite{1903.01382}. See \cite{1809.00442, 1809.06348, 1812.03136}, as well as the recent review \cite{1905.00287}, for more discussions.

Incidentally, the method used in Sec.(\ref{2}) can be used to compute GUP correction to the Schwinger effect as well. Since the steps are nearly identical, we only state the result here: 
the pair production rate goes like 
\begin{equation}
\Gamma_\text{GUP-dS} = \exp\left[-\left(\frac{\pi \hbar m_e^2}{eE} - \alpha \pi^3  e E \right)\right],
\end{equation}
which agrees -- up to a constant numerical factor in the second term linear in $E$ -- with the the result in \cite{1310.6966} obtained using a more rigorous method\footnote{Again, modulo the term that corresponds to the one-loop effect discussed in Footnote 2.}. This gives another support to the validity of our heuristic method. (However, to be fair, it is inconsistent with \cite{MWY}, in which the sign of the second term is opposite, although both \cite{1310.6966} and \cite{MWY} involve a positive GUP parameter.)

\section{Discussion: Can Heuristic Arguments Be Trusted?}\label{4}

Schwinger particle production by external electric field can be heuristically derived using the Heisenberg's uncertainty principle \cite{textbook}.
In this work, we provided a heuristic derivation of the Schwinger effect in anti-de Sitter spacetime (similarly for the de Sitter case) using the so-called extended uncertainty principle (EUP). 
We found that in order to obtain the known correct result, the EUP parameter must be negative in AdS spacetime, and positive in dS spacetime. 

We have further discussed why using the known result for Schwinger pair production rate to determine the sign of EUP parameter is more reasonable than using Hawking radiation, though both derivations are heuristic. Essentially, this is because Hawking temperature takes different forms depending on the curvature ($k$) of the spatial sections of the topological black hole spacetime, and while there are arguments that EUP depends on either the geometry or the topology of the underlying manifold, it is still not clear which argument is correct. On the other hand, the Schwinger effect does not depend on $k$ (see the Appendix), so this sidestepped the problem.

Nevertheless, the sign of EUP parameter -- like that of GUP -- still requires further studies, as different considerations and methods seem to yield different results. This issue requires a better understanding so that EGUP can be better employed as a phenomenological tool for us to investigate the interface of quantum mechanics and gravity. 

In fact, a greater issue is at hand: the GUP/EUP literature is full of heuristic arguments, the results of which are not all consistent. Our result has further highlighted this. Unfortunately it is far from clear which heuristic derivative can be trusted, and which cannot. Though as far as the EUP is concerned, various derivations do agree on its \emph{form}, they do not quite agree with the \emph{sign} of the EUP parameter. Perhaps they all have certain merits, but the subtleties are yet to be fully understood.

Finally, let us remark that this is not the only problem with EUP and GUP -- the ``heuristic'' treatment permeates much deeper throughout the literature. In the standard Heisenberg uncertainty principle, it is quite clear what $\Delta x^i$ and $\Delta p_i$ means. In Euclidean space, the Cartesian coordinates $\{x^i\}$ cover the whole space. We can interpret $p_{i}$ as the projections of the physical (conserved) momentum vector onto the global Cartesian axes in physical space. For EUP and GUP, on the other hand, which typically deal with spacetimes with nontrivial curvature, it is no longer clear what $\{x^i, p_j\}$ even mean. In particular, since physical distances on a curved manifold is usually not the same as coordinate distances, the associated momenta to the coordinates $\{x^i\}$ may not be the same as the physical momenta either.

Generalizations of the uncertainty principle are sometimes carried out at the level of the Heisenberg algebra. There is a vast literature about this, see, for example, \cite{1611.00001} and the references therein. However, the canonical Heisenberg algebra is simply the global shift-isometry algebra for Euclidean space, expressed in terms of Cartesian coordinates. This means that one can write a vector as a decomposition of the coordinate basis: $\vec{x} = x\vec{i} + y\vec{j} + z\vec{k}$, and similarly $\vec{p} = p_x\vec{i} + p_y\vec{j} + p_z\vec{k}$. This is often implicitly assumed even when the Heisenberg algebra has been modified, which appears to be (potentially) problematic, unless the underlying geometry admits a global coordinate chart (intrinsically flat).

This is not just being pedantic about the mathematics. Physics is ultimately about testing some predictions or observations. How then should we make testable prediction if the physical meaning of the underlying coordinates is not even clear? However, the required level of rigor to understand GUP and EUP correctly is as yet unavailable, since it is essentially a subject under development. For more discussions and open problems, see the review by Hossenfelder \cite{1203.6191}. This means that it is even more important to further understand whether heuristic treatments can be trusted, as the present work cast even more doubt into what the correct EUP should be.

\section*{Appendix: Euclidean Derivation of the Schwinger Effect in AdS Spacetime}

First, let us review the Euclidean method that allows us to compute Schwinger pair production rate in Minkowski spacetime. Upon Wick rotation $t \mapsto \tau = it$, Minkowski space (now Euclidean space) in the $\tau$-$r$ plane can be written in the polar form (the problem is essentially 2-dimensional): 
\begin{equation}
\text{d} s^2 = \text{d} R^2 + R^2 \text{d} \psi^2. 
\end{equation}
One construct an effective action  $S_\text{eff}=m\ell-eEA$, where $\ell$ and $A$ are the length (circumference) and the area of a circle of radius $R$ centered at an arbitrary fixed point. 
(We prefer not to include $\hbar$ in this effective action because it is a quantity constructed from classical geometry; the actual action is then $S=S_\text{eff}/\hbar$, which is rightfully dimensionless.)
Then
\begin{equation}
S_\text{eff}=mL-eEA= 2\pi m R - \pi q ER^2.
\end{equation}
Solving $\partial S_\text{eff}/\partial R =0 $ gives the extremal value $R_\text{ext}=m/(eE)$, which upon substituting back into the action gives $S_\text{eff}=\pi m^2/(eE)$. The pair production rate is $\exp(-S_\text{eff}/\hbar)$. The method is well-known, and was mentioned in, e.g., \cite{1411.1787, 0110178v3}. 

This method is readily generalized to anti-de Sitter spacetime. Circumference and area are best computed in the generalized version of polar coordinates -- the geodesic polar coordinates. Around an arbitrary fixed point, the metric of a space of constant negative Gaussian curvature $K$ in 2-dimensions has the following form (see Corollary 7.2.1 of \cite{kuhnel}):
\begin{equation}\label{k}
\text{d}s^2 = \text{d}r^2 + \frac{1}{-K}\sinh^2(\sqrt{-K}r) \text{d}\psi^2.
\end{equation}
Given the 2-dimensional AdS metric in static coordinates,
\begin{equation}
\text{d}s^2 =\left(1+\frac{\tilde{r}^2}{L^2}\right)\text{d}\tau^2+\left(1+\frac{\tilde{r}^2}{L^2}\right)^{-1}\text{d}\tilde{r}^2,
\end{equation}
we have $K=-1$, and so 
\begin{equation}
\text{d}s^2 = \text{d}r^2 + L^2\sinh^2(r/L) \text{d}\psi^2.
\end{equation}
The effective action $S=m\ell-eEA$ has circumference
\begin{equation}
\ell=\int_0^{2\pi} L \sinh(R/L) ~\text{d} \psi = 2\pi L \sinh(R/L),
\end{equation}
and area
\begin{equation}
A=\int_0^{2\pi} \int_0^{R} L \sinh(r/L) ~\text{d} r\text{d} \psi = 2\pi L^2[\cosh(R/L)-1].
\end{equation}
Solving for $\partial S_\text{eff}/\partial R =0 $ gives the extremal value 
\begin{equation}
R_\text{ext}=\frac{L}{2}\ln\left(\frac{eEL+m}{eEL-m}\right).
\end{equation}
Consequently,
\begin{equation}
S_\text{eff-AdS}=2\pi L^2\left(eE-\sqrt{e^2E^2-\frac{m^2}{L^2}}\right).
\end{equation}
This calculation only depends on the Gaussian curvature of the Euclidean manifold. It can be shown that AdS metric with different foliations such that
\begin{equation}
\text{d}s^2 =\left(k+\frac{\tilde{r}^2}{L^2}\right)\text{d}\tau^2+\left(k+\frac{\tilde{r}^2}{L^2}\right)^{-1}\text{d}\tilde{r}^2,
\end{equation}
also gives $K=-1/L^2$, and so the result is independent of $k$.

Note that alternatively, if we are only interested in the first few correction terms of the pair production rate, we can simply take a geodesic disk and compute with the well-known formula from differential geometry (see, e.g. Theorem 3.1 of \cite{grey}, which gives results for higher dimensions as well)
\begin{equation}
A=\pi R^2 \left(1-\frac{KR^2}{12}+\cdots \right),
\end{equation}
and
\begin{equation}
\ell=2\pi R\left(1-\frac{KR^2}{6}+\cdots \right),
\end{equation}
so that 
\begin{equation}
S_\text{eff-AdS} \approx 2\pi m R\left(1+\frac{R^2}{6L^2}\right)-\pi e E R^2 \left(1+\frac{R^2}{12 L^2}\right).
\end{equation}
Again we can solve for $\partial S_\text{eff}/\partial R =0 $ and subsitute the extremal value $R_\text{ext}$ into the effective action. This gives, after some cumbersome algebraic manipulations, the final result:
\begin{equation}
S_\text{eff-AdS} \approx \frac{\pi m^2}{eE} + \frac{1}{4}\frac{\pi m^4}{e^3E^3L^2}.
\end{equation}
Note that this method also does not depend on $k$.

\begin{acknowledgments}
YCO thanks the National Natural Science Foundation of China (No.11705162, No.11922508) and the Natural Science Foundation of Jiangsu Province (No.BK20170479) for funding support. He also thanks Brett McInnes for useful suggestions, and the referees for useful suggestions on improving the manuscript.
\end{acknowledgments}


\begin{thebibliography}{99}

\bibitem{pc}
Pisin Chen, Claudio Pellegrini, ``Boiling the Vacuum with Ultra Intense Lasers”, in \emph{Quantum Aspects of Beam Physics}, P. Chen, ed. p.571, World Scientific (1999).

\bibitem{Schwinger} Julian Schwinger, ``On Gauge Invariance and Vacuum Polarization", {\hypersetup{urlcolor=vividviolet}\href{https://journals.aps.org/pr/abstract/10.1103/PhysRev.82.664}{Phys. Rev. \textbf{82} (1951) 664}}.

\bibitem{textbook} Viatcheslav Mukhanov, Sergei Winitzki, \emph{Introduction to Quantum Effects in Gravity}, p.11, Cambridge University Press (2007).

\bibitem{0005078}
Sang Pyo Kim. Don N. Page, ``Schwinger Pair Production via Instantons in Strong Electric Fields'', {\hypersetup{urlcolor=vividviolet}\href{https://journals.aps.org/prd/abstract/10.1103/PhysRevD.65.105002}{Phys.R ev. D \textbf{65} (2002) 105002}}, \href{https://arxiv.org/abs/hep-th/0005078}{[arXiv:hep-th/0005078]}.

\bibitem{0501169}
Boris Pioline, Jan Troost, ``Schwinger Pair Production in $\text{AdS}_2$'', {\hypersetup{urlcolor=vividviolet}\href{https://iopscience.iop.org/article/10.1088/1126-6708/2005/03/043}{JHEP \textbf{0503} (2005) 043}}, \href{https://arxiv.org/abs/hep-th/0501169}{[arXiv:hep-th/0501169]}.

\bibitem{0803.2555}
Sang Pyo Kim, Don N. Page, ``Schwinger Pair Production in $\text{dS}_2$ and $\text{AdS}_2$'', {\hypersetup{urlcolor=vividviolet}\href{https://journals.aps.org/prd/abstract/10.1103/PhysRevD.78.103517}{Phys. Rev. D \textbf{78} (2008) 103517}}, \href{https://arxiv.org/abs/0803.2555}{[arXiv:0803.2555 [hep-th]]}. 

\bibitem{1407.4569}
Rong-Gen Cai, Sang Pyo Kim, ``One-Loop Effective Action and Schwinger Effect in (Anti-)de Sitter Space'', {\hypersetup{urlcolor=vividviolet}\href{https://link.springer.com/article/10.1007\%2FJHEP09\%282014\%29072}{JHEP \textbf{09} (2014) 072}}, \href{https://arxiv.org/abs/1407.4569v3}{[arXiv:1407.4569 [hep-th]]}.

\bibitem{BF}
Peter Breitenlohner, Daniel Z. Freedman, ``Positive Energy in Anti-de Sitter Backgrounds and Gauged Extended Supergravity'', {\hypersetup{urlcolor=vividviolet}\href{https://www.sciencedirect.com/science/article/abs/pii/0370269382906438?via\%3Dihub}{Phys. Lett. B \textbf{115} (1982) 197}}.

\bibitem{1804.04140}
Prasant Samantray, Suprit Singh, ``Schwinger Pair Production in Hot Anti-de Sitter Space'', {\hypersetup{urlcolor=vividviolet}\href{https://journals.aps.org/prd/abstract/10.1103/PhysRevD.99.085006}{Phys. Rev. D \textbf{99} (2019) 085006}}, \href{https://arxiv.org/abs/1804.04140}{[arXiv:1804.04140 [hep-th]]}.

\bibitem{0204110}
Sayed Fawad Hassan, Martin S. Sloth, ``Trans-Planckian Effects in Inflationary Cosmology and the Modified Uncertainty Principle'', {\hypersetup{urlcolor=vividviolet}\href{https://www.sciencedirect.com/science/article/pii/S0550321303007909?via\%3Dihub}{Nucl. Phys. B \textbf{674} (2003) 434}}, \href{https://arxiv.org/abs/hep-th/0204110}{[arXiv:hep-th/0204110]}.

\bibitem{0709.2307}
Mu-in Park, ``The Generalized Uncertainty Principle in (A)dS Space and the Modification of Hawking Temperature from the Minimal Length'', {\hypersetup{urlcolor=vividviolet}\href{https://www.sciencedirect.com/science/article/pii/S0370269307014980?via\%3Dihub}{Phys. Lett. B \textbf{659} (2008) 698}}, \href{https://arxiv.org/abs/0709.2307}{[arXiv:0709.2307 [hep-th]]}.

\bibitem{0411086v1}
Brett Bolen, Marco Cavaglia, ``(Anti-)de Sitter Black Hole Thermodynamics and the Generalized Uncertainty Principle'', {\hypersetup{urlcolor=vividviolet}\href{https://link.springer.com/article/10.1007/s10714-005-0108-x}{Gen. Rel. Grav. \textbf{37} (2005) 1255}}, \href{https://arxiv.org/abs/gr-qc/0411086v1}{[arXiv:gr-qc/0411086]}.



\bibitem{0106080}
Ronald J. Adler, Pisin Chen, David I. Santiago, ``The Generalized Uncertainty Principle and Black Hole Remnants'', {\hypersetup{urlcolor=vividviolet}\href{https://link.springer.com/10.1023/A:1015281430411}{Gen. Rel. Grav. \textbf{33} (2001) 2101}}, \href{https://arxiv.org/abs/gr-qc/0106080}{[arXiv:gr-qc/0106080]}.

\bibitem{1511.08221}
Steven B. Giddings, ``Hawking Radiation, the Stefan-Boltzmann law, and Unitarization'', {\hypersetup{urlcolor=vividviolet}\href{https://www.sciencedirect.com/science/article/pii/S0370269316000022?via\%3Dihub}{Phys. Lett. B \textbf{754} (2016) 39}}, \href{https://arxiv.org/abs/1511.08221}{[arXiv:1511.08221 [hep-th]]}.



\bibitem{Bambi}
Cosimo Bambi, Federico R. Urban, ``Natural Extension of the Generalised Uncertainty Principle'', {\hypersetup{urlcolor=vividviolet}\href{http://iopscience.iop.org/article/10.1088/0264-9381/25/9/095006/meta}{Class. Quant. Grav. \textbf{25} (2008) 095006}}, \href{https://arxiv.org/abs/0709.1965}{[arXiv:0709.1965 [gr-qc]]}.

\bibitem{1812.10045}
Matthew J. Lake, Marek Miller, Ray F. Ganardi, Zheng Liu, Shi-Dong Liang, Tomasz Paterek, ``Generalised Uncertainty Relations From Superpositions of Geometries'', {\hypersetup{urlcolor=vividviolet}\href{https://iopscience.iop.org/article/10.1088/1361-6382/ab2160}{Class. Quant. Grav. \textbf{36} (2019) 155012}}, \href{https://arxiv.org/abs/1812.10045}{[arXiv:1812.10045 [quant-ph]]}.

\bibitem{1110.0201}
Salvatore Mignemi, ``Classical and Quantum Mechanics of the Nonrelativistic Snyder Model in Curved Space'',  {\hypersetup{urlcolor=vividviolet}\href{https://iopscience.iop.org/article/10.1088/0264-9381/29/21/215019}{Class. Quant. Grav. \textbf{29} (2012) 215019}}, \href{https://arxiv.org/abs/1110.0201}{[arXiv:1110.0201 [hep-th]]}.

\bibitem{1911.08921}
Sebasti\'{a}n A. Franchino-Vi\~{n}as, Salvatore Mignemi, ``Asymptotic Freedom For $\lambda \phi^4_{\star}$ QFT In Snyder-de Sitter Space'', {\hypersetup{urlcolor=vividviolet}\href{https://link.springer.com/article/10.1140/epjc/s10052-020-7918-6}{Eur. Phys. J. C \textbf{80} (2020) 5, 382}}, \href{https://arxiv.org/abs/1911.08921}{[arXiv:1911.08921 [hep-th]]}.

\bibitem{F}
Raimundo N. Costa Filho, Jo\~{a}o P.M. Braga, Jorge H.S. Lira, José S. Andrade, ``Extended Uncertainty From First Principles'', {\hypersetup{urlcolor=vividviolet}\href{https://www.sciencedirect.com/science/article/pii/S0370269316001313?via\%3Dihub}{Phys. Lett. B \textbf{755} (2016) 367}}.

\bibitem{0306080}
Alexey Golovnev, Lev Vasil'evich Prokhorov, ``Uncertainty Relations in Curved Spaces'', 	{\hypersetup{urlcolor=vividviolet}\href{http://iopscience.iop.org/article/10.1088/0305-4470/37/7/017/meta}{J. Phys. A \textbf{37} (2004) 2765}}, \href{https://arxiv.org/abs/quant-ph/0306080}{[arXiv:quant-ph/0306080]}.

\bibitem{1804.02551}
Thomas Sch\"urmann, ``Uncertainty Principle on 3-Dimensional Manifolds of Constant Curvature'', {\hypersetup{urlcolor=vividviolet}\href{https://link.springer.com/article/10.1007\%2Fs10701-018-0173-0}{Found. Phys. \textbf{48} (2018) 716}}, \href{https://arxiv.org/abs/1804.02551}{[arXiv:1804.02551 [quant-ph]]}.

\bibitem{Mignemi}
Salvatore Mignemi, ``Extended Uncertainty Principle and the Geometry of (Anti)-de Sitter Space'', {\hypersetup{urlcolor=vividviolet}\href{https://www.worldscientific.com/doi/abs/10.1142/S0217732310033426}{Mod. Phys. Lett. A \textbf{25} (2010) 1697}}, \href{https://arxiv.org/abs/0909.1202v2}{[arXiv:0909.1202 [gr-qc]]}.

\bibitem{0911.5695}
Subir Ghosh, Salvatore Mignemi, ``Quantum Mechanics in de Sitter Space'', {\hypersetup{urlcolor=vividviolet}\href{https://link.springer.com/article/10.1007\%2Fs10773-011-0692-3}{Int. J. Theor. Phys. \textbf{50} (2011) 1803}}, \href{https://arxiv.org/abs/0911.5695}{[arXiv:0911.5695 [hep-th]]}.

\bibitem{9405067}
Achim Kempf, ``Quantum Field Theory with Nonzero Minimal Uncertainties in Positions and Momenta'', {\hypersetup{urlcolor=vividviolet}\href{https://link.springer.com/article/10.1007\%2FBF01690456}{Czechoslovak Journal of Physics \textbf{44} (1994) 1041}}, \href{https://arxiv.org/abs/hep-th/9405067}{[arXiv:hep-th/9405067]}.

\bibitem{judge1}
 D. Judge, ``On the Uncertainty Relation for $L_z$ and $\phi$'', {\hypersetup{urlcolor=vividviolet}\href{https://linkinghub.elsevier.com/retrieve/pii/S0375960163962832}{Phys. Lett. \textbf{5} (1963) 189}}.

\bibitem{judge2}
D. Judge, ``On the Uncertainty Relation for Angle Variables'', {\hypersetup{urlcolor=vividviolet}\href{https://link.springer.com/article/10.1007/BF02733639}{II Nuovo Cimento \textbf{31 (2)} (1964) 332}}.

\bibitem{1512.05716}
Adam R. Brown, ``Schwinger Pair Production at Nonzero Temperatures or in Compact Directions'', {\hypersetup{urlcolor=vividviolet}\href{https://journals.aps.org/prd/abstract/10.1103/PhysRevD.98.036008}{Phys. Rev. D \textbf{98} (2018) 036008}}, \href{https://arxiv.org/abs/1512.05716}{[arXiv:1512.05716 [hep-th]]}.

\bibitem{1912.07094}
Matthew J. Lake, Marek Miller, Shi-Dong Liang, ``Generalised Uncertainty Relations For Angular Momentum and Spin in Quantum Geometry'', {\hypersetup{urlcolor=vividviolet}\href{https://www.mdpi.com/2218-1997/6/4/56}{Universe \textbf{6} (2020) 56}}, \href{https://arxiv.org/abs/1912.07094}{[arXiv:1912.07094 [gr-qc]]}.

\bibitem{9808032}
Danny Birmingham, ``Topological Black Holes in Anti-de Sitter Space'', {\hypersetup{urlcolor=vividviolet}\href{https://iopscience.iop.org/article/10.1088/0264-9381/16/4/009}{Class. Quant. Grav. \textbf{16} (1999) 1197}}, \href{https://arxiv.org/abs/hep-th/9808032}{[arXiv:hep-th/9808032]}.


\bibitem{0607010}
Fabio Scardigli, ``Hawking Temperature for Various Kinds of Black Holes From Heisenberg Uncertainty Principle'', \href{https://arxiv.org/abs/gr-qc/0607010}{[arXiv:gr-qc/0607010]}.

\bibitem{Hamil}
B. Hamil, M. Merad, ``Schwinger Mechanism on de Sitter Background'', {\hypersetup{urlcolor=vividviolet}\href{https://www.worldscientific.com/doi/abs/10.1142/S0217751X18501774}{Int. J. Mod. Phys. A \textbf{33} (2018) 1850177}}.

\bibitem{1401.4137}
Markus B. Fr\"ob, Jaume Garriga, Sugumi Kanno, Misao Sasaki, Jiro Soda, Takahiro Tanaka, Alexander Vilenkin, ``Schwinger Effect in de Sitter Space'', {\hypersetup{urlcolor=vividviolet}\href{https://iopscience.iop.org/article/10.1088/1475-7516/2014/04/009}{JCAP \textbf{1404} (2014) 009}}, \href{https://arxiv.org/abs/1401.4137}{[arXiv:1401.4137 [hep-th]]}.

\bibitem{1411.1787}
Willy Fischler, Phuc H. Nguyen, Juan F. Pedraza, Walter Tangarife, ``Holographic Schwinger Effect In de Sitter Space'', {\hypersetup{urlcolor=vividviolet}\href{https://journals.aps.org/prd/abstract/10.1103/PhysRevD.91.086015}{Phys. Rev. D 91 (2015) 086015}}, \href{https://arxiv.org/abs/1411.1787}{[arXiv:1411.1787 [hep-th]]}.


\bibitem{9311147}
 Achim Kempf, ``Uncertainty Relation in Quantum Mechanics with Quantum Group Symmetry'', {\hypersetup{urlcolor=vividviolet}\href{https://aip.scitation.org/doi/10.1063/1.530798}{J. Math. Phys. \textbf{35} (1994) 4483}}, \href{https://arxiv.org/abs/hep-th/9311147}{[arXiv:hep-th/9311147]}.

\bibitem{9412167}
Achim Kempf, Gianpiero Mangano, Robert B. Mann, ``Hilbert Space Representation of the Minimal Length Uncertainty Relation'', {\hypersetup{urlcolor=vividviolet}\href{https://journals.aps.org/prd/abstract/10.1103/PhysRevD.52.1108}{Phys. Rev. D \textbf{52} (1995) 1108}}, \href{https://arxiv.org/abs/hep-th/9412167}{[arXiv:hep-th/9412167]}.

\bibitem{9301067}
Michele Maggiore, ``A Generalized Uncertainty Principle in Quantum Gravity'',  {\hypersetup{urlcolor=vividviolet}\href{https://www.sciencedirect.com/science/article/pii/0370269393914018?via\%3Dihub}{Phys. Lett. B \textbf{304} (1993) 65}}, \href{https://arxiv.org/abs/hep-th/9301067}{[arXiv:hep-th/9301067]}.

\bibitem{9305163}
Michele Maggiore, ``Quantum Groups, Gravity, and the Generalized Uncertainty Principle'', {\hypersetup{urlcolor=vividviolet}\href{https://journals.aps.org/prd/abstract/10.1103/PhysRevD.49.5182}{Phys. Rev. D \textbf{49} (1994) 5182}}, \href{https://arxiv.org/abs/hep-th/9305163}{[arXiv:hep-th/9305163]}.

\bibitem{9904025}
Fabio Scardigli, ``Generalized Uncertainty Principle in Quantum Gravity from Micro-Black Hole Gedanken Experiment'', {\hypersetup{urlcolor=vividviolet}\href{https://www.sciencedirect.com/science/article/pii/S0370269399001677?via\%3Dihub}{Phys. Lett. B \textbf{452} (1999) 39}}, \href{https://arxiv.org/abs/hep-th/9904025}{[arXiv:hep-th/9904025]}.

\bibitem{9904026} 
Ronald J. Adler, David I. Santiago, ``On Gravity and the Uncertainty Principle'', {\hypersetup{urlcolor=vividviolet}\href{https://www.worldscientific.com/doi/abs/10.1142/S0217732399001462}{Mod. Phys. Lett. A \textbf{14} (1999) 1371}}, \href{https://arxiv.org/abs/gr-qc/9904026}{[arXiv:gr-qc/9904026]}. 

\bibitem{6} David J. Gross, Paul F. Mende, ``String Theory Beyond the Planck Scale'', {\hypersetup{urlcolor=vividviolet}\href{https://www.sciencedirect.com/science/article/pii/0550321388903902}{Nucl. Phys. B \textbf{303} (1988) 407}}.

\bibitem{7} Daniele Amati, Marcello Ciafolini, Gabriele Veneziano, ``Can Spacetime be Probed Below the String Size?'', {\hypersetup{urlcolor=vividviolet}\href{https://www.sciencedirect.com/science/article/pii/037026938991366X}{Phys. Lett. B \textbf{216} (1989) 41}}.

\bibitem{8} Kenichi Konishi, Giampiero Paffuti, Paolo Provero , ``Minimum Physical Length and the Generalized Uncertainty Principle in String Theory'', {\hypersetup{urlcolor=vividviolet}\href{https://www.sciencedirect.com/science/article/pii/0370269390919274}{Phys. Lett. B \textbf{234} (1990) 276}}.

\bibitem{1001.1205}
Ronald J. Adler, ``Six Easy Roads to the Planck Scale'', {\hypersetup{urlcolor=vividviolet}\href{https://aapt.scitation.org/doi/10.1119/1.3439650}{Am. J. Phys. \textbf{78} (2010) 925}}, \href{https://arxiv.org/abs/1001.1205}{[arXiv:1001.1205 [gr-qc]]}.

\bibitem{0912.2253}
Petr Jizba, Hagen Kleinert, Fabio Scardigli, ``Uncertainty Relation on World Crystal and its Applications to Micro Black Holes'', {\hypersetup{urlcolor=vividviolet}\href{https://journals.aps.org/prd/abstract/10.1103/PhysRevD.81.084030}{Phys. Rev. D \textbf{81} (2010) 084030}}, \href{https://arxiv.org/abs/0912.2253v2}{[arXiv:0912.2253 [hep-th]]}.

\bibitem{1407.0113}
Fabio Scardigli, Roberto Casadio, “Gravitational tests of the Generalized Uncertainty Principle”, {\hypersetup{urlcolor=vividviolet}\href{https://link.springer.com/article/10.1007\%2FJHEP07\%282015\%29052}{Eur. Phys. J. C \textbf{75} (2015) 425}}, \href{https://arxiv.org/abs/1504.07637}{[arXiv:1407.0113 [hep-th]]}.

\bibitem{KLVY}
T. Kanazawa, G. Lambiase, G. Vilasi, A. Yoshioka, ``Noncommutative Schwarzschild Geometry and Generalized Uncertainty Principle'', {\hypersetup{urlcolor=vividviolet}\href{https://link.springer.com/article/10.1140\%2Fepjc\%2Fs10052-019-6610-1}{Eur. Phys. J. C \textbf{79} (2019) 95}}.

\bibitem{1903.01382}
Luca Buoninfante, Giuseppe Gaetano Luciano, Luciano Petruzziello, ``Generalized Uncertainty Principle and Corpuscular Gravity'', {\hypersetup{urlcolor=vividviolet}\href{https://link.springer.com/article/10.1140\%2Fepjc\%2Fs10052-019-7164-y}{Eur. Phys. J. C \textbf{79} (2019) 663}}, \href{https://arxiv.org/abs/1903.01382}{[arXiv:1903.01382 [gr-qc]]}.

\bibitem{1809.00442}
Yen Chin Ong, ``GUP-Corrected Black Hole Thermodynamics and the Maximum Force Conjecture", {\hypersetup{urlcolor=vividviolet}\href{https://linkinghub.elsevier.com/retrieve/pii/S0370269318306828}{Phys. Lett. B \textbf{785} (2018) 217}}, \href{https://arxiv.org/abs/1809.00442}{[arXiv:1809.00442 [gr-qc]]}.

\bibitem{1809.06348}
Yen Chin Ong, Yuan Yao, ``Generalized Uncertainty Principle and White Dwarfs Redux: How Cosmological Constant Protects Chandrasekhar Limit", {\hypersetup{urlcolor=vividviolet}\href{https://journals.aps.org/prd/abstract/10.1103/PhysRevD.98.126018}{Phys. Rev. D \textbf{98} (2018) 12018}}, \href{https://arxiv.org/abs/1809.06348}{[arXiv:1809.06348 [gr-qc]]}.

\bibitem{1812.03136}
Yuan Yao, Meng-Shi Hou, Yen Chin Ong, ``A Complementary Third Law for Black Hole Thermodynamics'', {\hypersetup{urlcolor=vividviolet}\href{https://link.springer.com/article/10.1140\%2Fepjc\%2Fs10052-019-7003-1}{Eur. Phys. J. C \textbf{79} (2019) 513}}, \href{https://arxiv.org/abs/1812.03136}{[arXiv:1812.03136 [gr-qc]]}.

\bibitem{1905.00287}
Fabio Scardigli, ``The Deformation Parameter of the Generalized Uncertainty Principle'', {\hypersetup{urlcolor=vividviolet}\href{https://iopscience.iop.org/article/10.1088/1742-6596/1275/1/012004}{J. Phys. Conf. Ser. \textbf{1275} (2019) 012004}}, \href{https://arxiv.org/abs/1905.00287}{[arXiv:1905.00287 [hep-th]]}.'

\bibitem{1310.6966}
Salah Haouat, Khireddine Nouicer, ``Influence of a Minimal Length on the Creation of Scalar Particles'', {\hypersetup{urlcolor=vividviolet}\href{https://journals.aps.org/prd/abstract/10.1103/PhysRevD.89.105030}{Phys. Rev. D \textbf{89} (2014) 105030}}, \href{https://arxiv.org/abs/1310.6966}{[arXiv:1310.6966 [hep-th]]}.

\bibitem{MWY} Ben-Rong Mu, Peng Wang, Hai-Tang Yang, ``Minimal Length Effects on Schwinger Mechanism'', {\hypersetup{urlcolor=vividviolet}\href{https://doi.org/10.1088/0253-6102/63/6/715}{Commun. Theor. Phys. \textbf{63} (2015) 715}}, \href{https://arxiv.org/abs/1501.06020}{[arXiv:1501.06020 [gr-qc]]}.

\bibitem{1611.00001}
Syed Masood, Mir Faizal, Zaid Zaz, Ahmed Farag Ali, Jamil Raza, Mushtaq B Shah, ``The Most General Form of Deformation of the Heisenberg Algebra from the Generalized Uncertainty Principle'', {\hypersetup{urlcolor=vividviolet}\href{https://www.sciencedirect.com/science/article/pii/S0370269316306189?via\%3Dihub}{Phys. Lett. B \textbf{763} (2016) 218}}, \href{https://arxiv.org/abs/1611.00001}{[arXiv:1611.00001 [hep-th]]}.

\bibitem{1203.6191}
Sabine Hossenfelder, ``Minimal Length Scale Scenarios for Quantum Gravity'', {\hypersetup{urlcolor=vividviolet}\href{https://link.springer.com/article/10.12942/lrr-2013-2}{Living Rev. Relativity \textbf{16} (2013) 2}}, \href{https://arxiv.org/abs/1203.6191}{[arXiv:1203.6191 [gr-qc]]}.

\bibitem{0110178v3}
A. S. Gorsky, K. A. Saraikin, K. G. Selivanov, ``Schwinger Type Processes via Branes and Their Gravity Duals'', {\hypersetup{urlcolor=vividviolet}\href{https://linkinghub.elsevier.com/retrieve/pii/S0550321302000950}{Nucl. Phys. B \textbf{628} (2002) 270}}, \href{https://arxiv.org/abs/hep-th/0110178v3}{[arXiv:hep-th/0110178]}.

\bibitem{kuhnel}
Wolfgang K\"uhnel, \emph{Differential Geometry: Curves-Surfaces-Manifolds}, Second Ed., Student Mathematical Library Vol. 16, American Mathematical Society, 2006.

\bibitem{grey}
Alfred Gray, ``The Volume of a Small Geodesic Ball of a Riemannian Manifold'', {\hypersetup{urlcolor=vividviolet}\href{https://projecteuclid.org/euclid.mmj/1029001150}{Michigan Math. Jour. \textbf{20}  (1974) 329}}.

\end{thebibliography}
\end{document}